# Comparison of Image Scale Calibration Techniques:

# Known Pairs, Drift Scans and Aperture Grating.

Matthew B. James[1], Graeme L. White[2], Stephen G. Bosi[1], Rod R. Letchford[2]

1. University of New England, Armidale, NSW, Australia;
   m.b.james27@gmail.com; sbosi@une.edu.au
2. Centre for Astronomy, University of Southern Queensland, Toowoomba, QLD, Australia; graemewhiteau@gmail.com; Rod.Letchford@usq.edu.au

**Abstract**: We compared several techniques for calibrating angular separation between wide (>1 arcsec) pairs. These techniques are (i) reference pair calibration using α Cen AB orbital parameters, (ii) the video drift method, and (iii) the utilisation of an aperture diffraction grating with red filters of different passbands. Separations of 62 pairs were determined using these 3 calibration techniques and compared. It was found that α Cen AB and video drift methods are in good agreement.

A bias in the diffraction grating method of ~0.1 arcseconds in the separation of presented pairs revealed itself and could not be accounted for. We conclude that video drift calibration is probably the simplest and best way to undertake image scale calibration.

## 1. Introduction.

Wide double star astrometry consists of the measurement of the separation, $\rho$, and the position angle, PA, of pairs that are typically wider than ~1 arcsecond. Such astrometric measures are now obtained using "lucky imaging" and are the subject of many papers in the JDSO, and our two previous papers (James *et al.*, 2019a; James *et al.*, 2020a); hereafter Paper 1 and Paper 2. A detailed account of this work is also available in James (2019b).

The uncertainty in $\rho$ and PA measures is now the combination of the uncertainty in the calibration of the instrument, and the uncertainty in the individual *x-y* measurements of the star images on the detector. Repeated measurement of the *x-y* positions will yield the uncertainty in the measurements themselves, and a numerical value of that uncertainty is usually expressed as the Standard Error in the Mean, SEM, of the ensemble of *xy* measures.

The determination of the calibration/image scale of the detector, and the uncertainty in that image scale, is the subject of this paper.

We concentrate on the calibration process used for the separations and PA reported in our Paper 1 and Paper 2, and explore the accuracy and convenience of undertaking image scale calibration utilising a full aperture diffraction grating.

## 2. Methods of Determining Image Scale Calibration.

### 2.1 Reference Pair Calibration using α Cen AB Orbital Parameters.

In brief, reference pair calibration uses well-established positions of pairs to predict the ρ and PA at the epoch of observation.

For our Paper 1, the positions of the pair α Cen AB (Rigil Kentaurus and Toliman; $\rho \approx 5$ arcsec, PA ≈ 338°, $m_V = 0.01$, $m_V = 1.33$) were used for image scale calibration. α Cen AB has a well determined Grade 2 orbit (White *et al.*, 2018; Pourbaix & Boffin, 2016; Table 1) that was used to predict the ρ and PA of the stars at the date of observation. These values were then used to calibrate the image scale of the detector in units of pixels per arcsecond (px/arcsec). Paper 1 reports the positions of 10 pairs with an accuracy of ≈ 50 milli-arcsec (mas) in ρ, and ≈ 0.15° in PA, relative to GAIA positions.

The α Cen AB method approach to calibration was utilised in Paper 2, and the measures reported there are the weighted mean of the α Cen AB method and the drift scan method. For the 62 pairs in Paper 2 the formal internal uncertainties for the α Cen AB method contribution are $\delta\rho = 94$ mas and $\delta PA = 0.097°$. The image scale was found to be $5.664 \pm 0.023$ px/arcsec and used in both Paper 1 and Paper 2.

*Table 1: Orbital elements of α Cen AB used to determine the position of the pair at epoch. From Pourbaix & Boffin (2016) – from the WDS 6th Orbit Catalog.*

| $P$ Orbital Period (Years) | $a$ Semi-major axis (degrees) | $i$ Inclination (degrees) | $\Omega$ Ascending Node (degrees) | $T$ Time of Periastron passage | $e$ Eccentricity | $\Omega$ Longitude of Periastron (degrees) |
|---|---|---|---|---|---|---|
| 79.97 | 17.66 | 79.32 | 204.75 | 1955.66 | 0.524 | 232.3 |
| ±0.013 | ±0.026 | ±0.044 | ±0.087 | ±0.014 | ±0.0011 | ±0.11 |

## 2.2 Video Drift Method.

The video drift method is as outlined by Nugent & Iverson (2011). The target object is place outside the east side of the FoV of the detector and telescope tracking is disabled. The image drifts across the sensor. Using the rotation rate of the Earth (15.041 arcsecond per second) and the declination of the object, an image scale can be computed from the time stamp on the image. The video drift method also provided the orientation of the camera as the image drift is east to west, perpendicular to the meridian of the object.

The video drift approach to calibration was utilised in Paper 2, and the measures reported there are the weighted mean of the α Cen AB method and the drift scan method. For the 62 pairs the contribution of the video drift method to the formal internal uncertainties are $\delta\rho = 72$ mas and $\delta PA = 0.046°$. The image scale utilised in Paper 2, and used in conjunction with the α Cen AB orbital parameter determined image scale, was $5.657 \pm 0.002$ px/arcsec as found with the video drift method.

## 2.3 Aperture Diffraction Grating Method.

Aperture diffraction grating image scale calibration is explained in Argyle (2004 – as is the use of a full aperture grating to estimate astrometric measures). Using a diffraction grating on the front of a telescope transforms the approximate point source of star light into a diffraction pattern, the nature of which depends on the separations and width of the slits of the grating, and the wavelength of the light through the system. The linear distance between the zeroth, and first order diffraction peaks on the sensor, and the angular dispersion of these peaks, is used to calibrate image scale.

### 2.3.1 Angular Separation of the Diffraction Orders.

The angular separation, $z$, of the zero and the first order is determined from Equation 1. Wavelength, $\lambda$, in this work is the mean wavelength of the observation in mm, $(l+d)$ is the periodicity of the grating in mm, $z$ is the angular separation between the zeroth and first order peaks in arcsec, and 206265 is the number of arcseconds in 1 radian.

$$z = \frac{206265\,\lambda}{l+d}$$

*Equation 1: The angular distance between the zero and first order of a diffraction pattern. From Argyle, 2004, p. 184.*

### 2.3.2 Fraunhofer Diffraction Modelling.

Fraunhofer diffraction modelling was conducted for comparison with the other calibration methods in this paper and as validation for our observations. The diffraction pattern is given by Equation 2 (Hecht, 2001).

Figure 1 shows the modelled diffraction pattern using the passband of the telescope optics (noted below) with the Hα filter ($\lambda \approx 656.3$ nm with a bandpass of 7 nm) in place.

In Equation 2, $a$ is the grating periodicity (m), $b$ is slit width (m), $k$ is the wave number, and $I_0$ is the normalisation irradiance constant. The equations for $\alpha$ and $\beta$ are shown directly below Equation 2.

$$\frac{I(\theta)}{I_0} = \left(\frac{\sin\beta}{\beta}\right)^2 \left(\frac{\sin(N\alpha)}{\sin\alpha}\right)^2$$

$$\beta = \left(\frac{kb}{2}\right)\sin\theta \quad \text{and} \quad \alpha = \left(\frac{ka}{2}\right)\sin\theta$$

*Equation 2: Fraunhofer equation for multiple slit diffraction. Hecht, p. 452.*

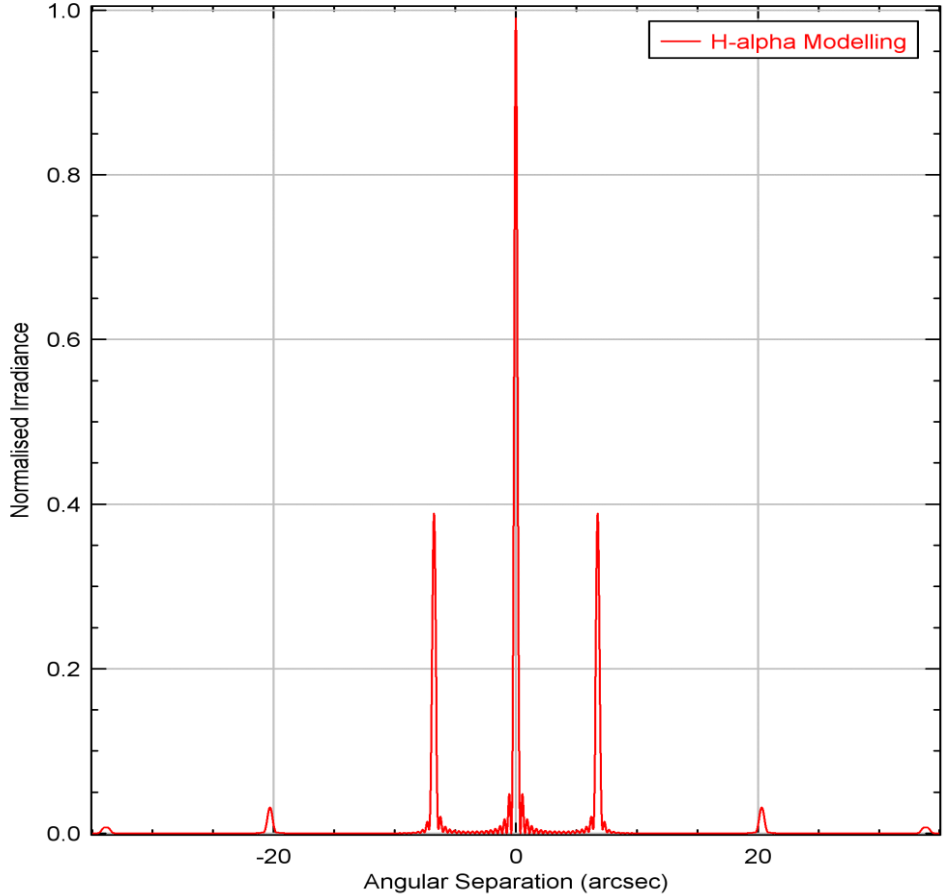

*Figure 1: Modelled diffraction pattern produced for the Webster Celestron C14, 18-slit aperture grating and Hα filter. The separation between orders is computed as 6.765 arcsec.*

Grating calibration technique is only useable with bright stars. The geometry of our diffraction grating blocks 50% of the light from entering the telescope and modelling shows that the first orders of the diffraction pattern are only ~40% the brightness of the zero order.

Some assumptions of the model presented are that (i) the distribution of intensity of wavelengths are uniform over the bandpass of the filter and (ii) that the 14-inch aperture of the telescope and central circular obstruction of the secondary mirror have no effect. The second assumption was modelled using Equation 3, and it was found to have a negligible effect on our precision, thus, modelled results in this work are of the effects of the 18-slit diffraction grating only (Equation 2).

The pronumerals in Equation 3 are the same as Equation 2 (described above), with the addition of, *c,* the is the ratio of the diameters of the central circular obstruction over the telescope's aperture and, $J_1$, a type 1 Bessel function.

$$\frac{I(\theta)}{I_0} = \frac{4}{(1-c^2)^2} \left[ \frac{J_1(ka\,sin\theta) - cJ_1(kac\,sin\theta)}{ka\,sin\theta} \right]^2$$

*Equation 3: Fraunhofer diffraction equation for circular aperture and the central circular obstruction of the secondary mirror. Hecht p. 469 and Goldberg & McCulloch, 1961.*

## 3. Observations Using the Full Aperture Grating Method.

### 3.1 Data Acquisition.

For the image scale comparison work in this paper, we introduce 60 diffraction grating calibration observations of 6 stars obtained over 3 nights of observations in ≈ 2019.9. Each star was observed 10 times; 5 observations with the grating oriented horizontally and 5 vertically relative to the approximate *x-y* orientation of the chip. This was a precaution to avoid potential bias due to asymmetry in the *x-y* scale in the focal plane of the telescope.

### 3.2 Equipment and Software.

The equipment and software used was the Bill Webster 14-inch Celestron Schmidt-Cassegrain telescope, ZWO ASI120MM-s camera, a custom-made eyepiece/camera flip box, and the software *SharpCaps* and *Reduc*. These are described in Paper 1 and James (2019b).

To this system was added a 3D printed aperture diffraction grating (Figure 2) with 18 parallel slats (there are 20 slats, but the 2 smaller edge slats do not continue the periodicity of the grating) with equal spacing and slat widths of $19.95 \pm 0.13$ mm. This uncertainty is incorporated into the final uncertainty value listed in Table 4.

The aperture diffraction grating was 3D printed with a Raise3D pro2 3D printer, owned and maintained by the University of New England (UNE). The 3D model was created using the *Sketchup* software, then passed on to *ideaMaker* as a *stereolithography* (STL) file for printing.

Observations of the 6 stars were made through three red filters:
- A Wratten #25 with extended red passband and a blue cut-off from 580 nm.
- A TR-TriColour red filter with a box-like transmission from 600 nm to 700 nm. (This is the red filter from a colour imaging filter set)
- A ZWO H$\alpha$ filter with 7 nm bandpass centred near 656.3 nm.

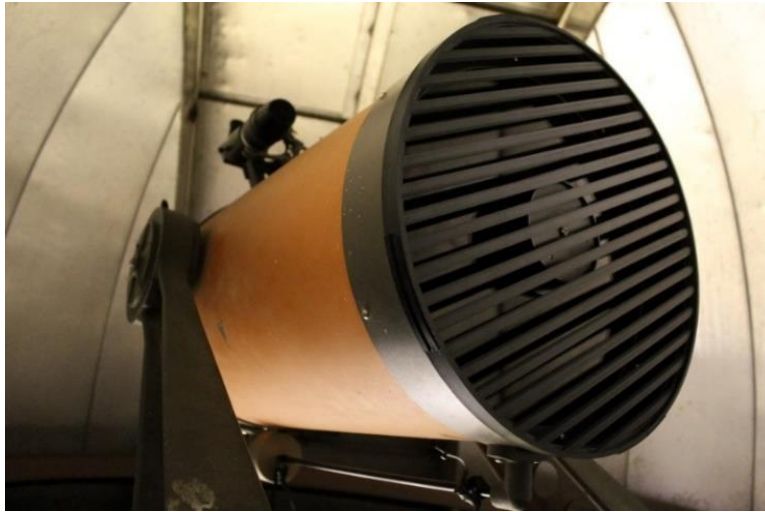

*Figure 2: The Webster 14-inch Schmidt-Cassegrain telescope of the Kirby observatory used for this work (and Papers 1 and 2, and James 2019b) with the 3D printed grating attached.*

### 3.3    Selection of Stars for the Full Aperture Grating.

The stars selected for diffraction calibration of the ZWO camera were single bright stars at mid-southern declination. A range of spectral types were chosen (Table 2). Sources for the values of (B-V) and Effective Temperatures are listed under the 'source' column. The majority of sources were found through the SIMBAD Astronomical Database.

*Table 2: Single stars selected for aperture grating observations.*

| Designation | ICRS Coordinate (J2000) | Spectral Type | B-V | $T_{eff}$ (K) | Source |
|---|---|---|---|---|---|
| HD 151804 | $16^h51^m34^s$-41°13'50" | O8I | 0.07 | 36660 | Morossi & Crivellari, 1980; Ducati, 2002 (VizieR) |
| o Sco | $16^h20^m38^s$-24°10'10" | A4II/III | 0.83 | 8128 | de Geus, de Zeeuw & Lub,1989; Ducati, 2002 (VizieR) |
| θ Sco A | $17^h37^m19^s$-42°59'52" | F1III | 0.40 | 7200 | Hohle, Neuhäuser & Schutz, 2010; UNSO, pg. 534 |
| $ω^2$ Sco | $16^h07^m24^s$-20°52'08" | G4II-III | 0.84 | 5200 | Franchini, Morossi & Malagnini, 1998; Ducati, 2002 (VizieR) |
| ε Sco | $16^h50^m10^s$-34°17'36" | K1III | 1.16 | 4560 | McWilliam, 1990; Ducati, 2002 (VizieR) |
| α Sco A | $16^h29^m25^s$-26°25'55" | M1.5I | 1.84 | 3660 | Ohanaka, *et al.*, 2013; Hoffleit & Warren, 1991 (VizieR) |

### 3.3    Mean Wavelength of the Star Light.

To utilise Equation 1, the wavelength of the light reaching the detector must be known. The mean wavelength of stellar light that is recorded by the detector is not just the product of the transmission of the filter, but includes the spectral response of each "system element" that has an effect on the mean wavelength of light detected by the camera. These system elements are:

- The star's black body curve (Table 2, column 'Source').
- The Earth's atmosphere (Berk, *et al.,* 2014 & 2015).
- The spectral reflectivity of the aluminised mirror coating (TYDEX).

- The spectral transmission of the soda glass correcting plate of the early C14 telescope (Abrisa),
- The spectral passband of the filters used;
    - The Wratten #25 filter (Peed, n.d.).
    - The TR TriColour RED filer (SIRIUS OPTICS).
    - The ZWO narrow bandpass Hα filter (ZWO).
- The spectral quantum efficiency of the camera chip (ON Semiconductor).

Figure 3 shows the spectral transmission of the air, the telescope, filter, and camera components. The *y*-axis is the relative transmission, the *x*-axis is the wavelength in nm, the red highlights are notable absorbance and the green dots is extrapolated absorbance.

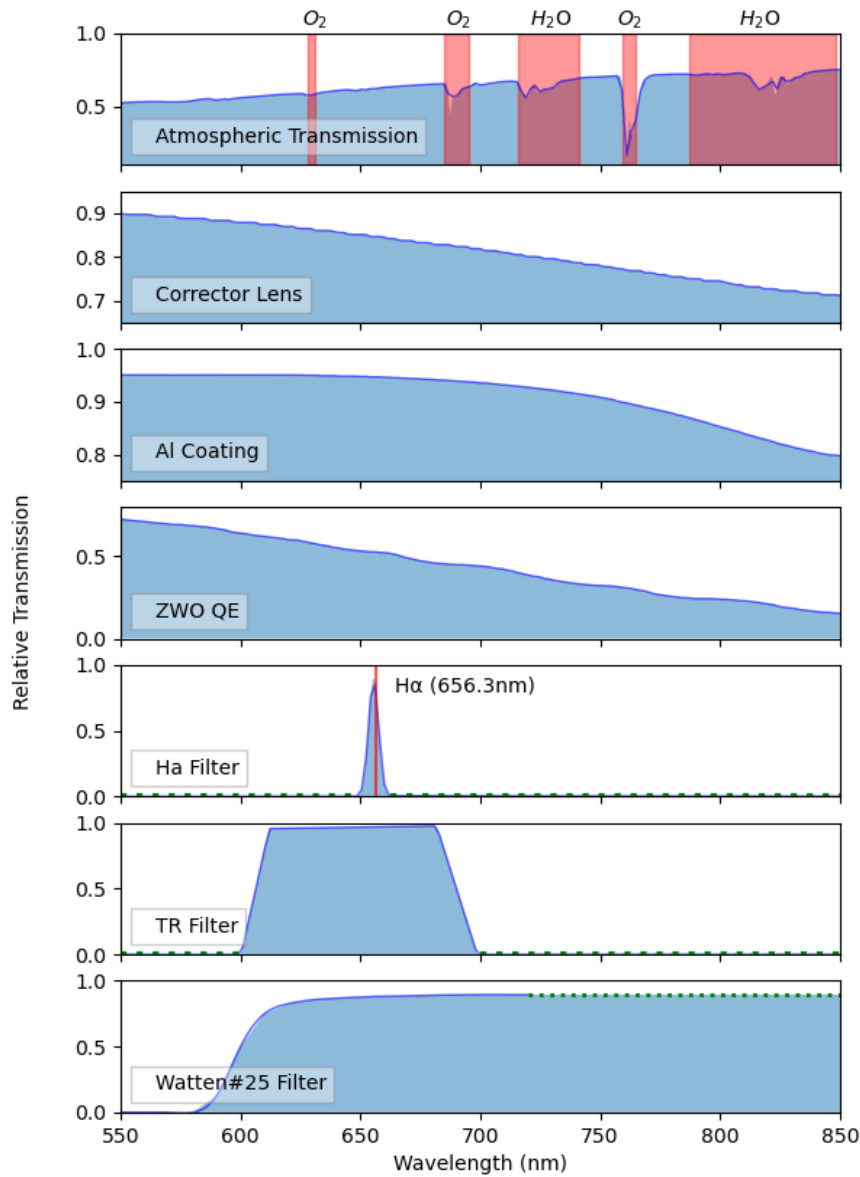

*Figure 3: Relative transmission plots of each system element; the air, telescope and camera components. The red highlights are notable absorbance and the green dots is extrapolated absorbance.*

## 4. Observations - The Image Scale from the Full Aperture Grating.

Figure 4 is the sum of ~2500 diffraction images of the bright star λ Sco co-added and normalised. These image frames were made with the Wratten #25, the TR TriColour filter, and the Hα filter. Measurement of diffraction peaks was undertaken using *Reduc* (treating the diffraction peaks as companion objects) as is done when analysing pairs normally.

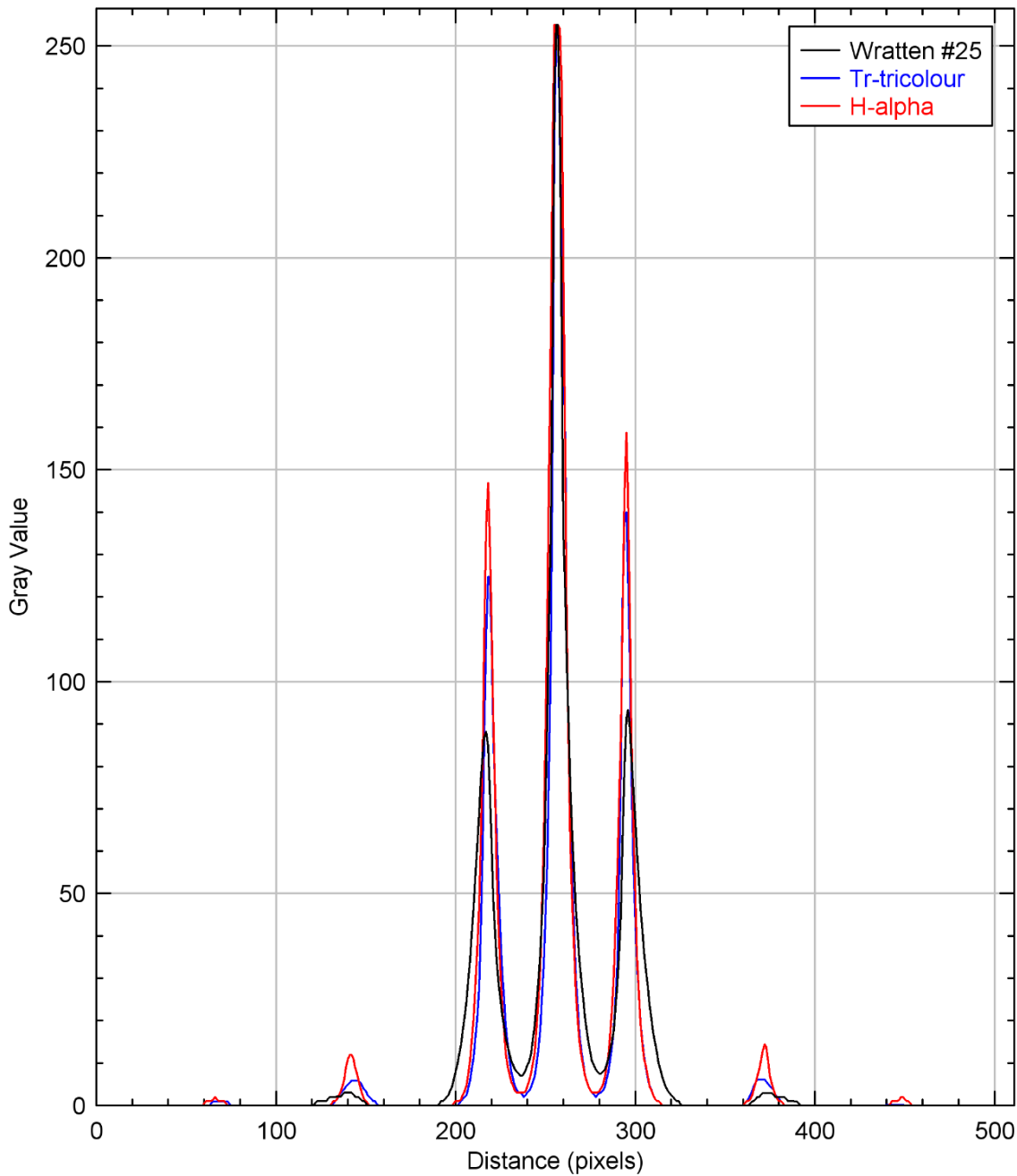

*Figure 4: Profile plot of λ Sco (B-type) diffraction pattern. Seen are the zero, first, second and third orders of the diffraction pattern. The theoretical version of this figure is Figure 1 (above).*

## 5. Precision, Errors and Bias of the Full Grating Filter.

The angular separation of the zero and first orders of the observed diffraction pattern, in arcseconds, and their *x-y* position on the sensor, in pixels, are used for image scale calibration. The uncertainty in the calibration arises from 3 main sources (i) the manufacturing precision of the grating, (ii) measurement of the linear *x-y* separation between the zero and first orders of the diffraction pattern across the chip, and (iii) the determination of the mean wavelength of the light for each filter.

The uncertainty for the grating spacing is the Standard Error of the Mean (SEM) of repeated micrometre measurement made along the lengths of each of the slats. The uncertainty in the *x-y* measurement of the peaks in the diffraction pattern is the uncertainty reported by the *reduc* software.

An uncertainty in the mean wavelength was difficult to determine as the spectral transmittance of many components of the optical system affected the final transmission profile (Figure 3). The approach was to estimate an approximate upper bound for an uncertainty given by the difference between the raw mean filter wavelength and the effective mean wavelength when all system components are accounted for.

The final uncertainty estimates presented in Table 3 are expressed as two components: the absolute random uncertainties of *x-y* separations expressed as an SEM (ran), and an estimate of the relative systematic uncertainties comprising the uncertainty estimate in the mean wavelength and the uncertainty in the manufacturing of the grating (slat widths and spacings) added in quadrature (sys).

One might also justify combining the systematic uncertainties by simple addition instead of quadrature because the wavelength uncertainty was based on an estimate of maximum change in mean wavelength rather than a statistical calculation. Simple addition results in a moderate increase in the reported systematic uncertainty (~30% for the broad-band filters and ~6% for Hα filter).

## 6. Results.

### 6.1 Grating Calibration and the Spectral Type.

Table 3 is a comparison of image scales and uncertainties determined with each of the 3 filters, and the B-V value of the observed star. As expected, precision increases with a narrower passband; the calculated image scale is more precise for the narrow band Hα filter (Table 4).

*Table 3: Image scale for each filter and stellar spectrum.*

| Name | B-V | Wratten #25 (px/arcsec) | Uncertainty (px/arcsec) | TR TriColour (px/arcsec) | Uncertainty (px/arcsec) | Hα (px/arcsec) | Uncertainty (px/ arcsec) |
|---|---|---|---|---|---|---|---|
| HD 151804 | 0.07 | 6.03 | ± 0.01(ran) ± 0.10(sys) | 5.731 | ± 0.003(ran) ± 0.054(sys) | 5.689 | ± 0.004(ran) ± 0.037(sys) |
| o Sco | 0.83 | 6.03 | ± 0.02(ran) ± 0.08(sys) | 5.741 | ± 0.008(ran) ± 0.042(sys) | 5.685 | ± 0.004(ran) ± 0.037(sys) |
| θ Sco | 0.40 | 6.02 | ± 0.01(ran) ± 0.07(sys) | 5.745 | ± 0.006(ran) ± 0.041(sys) | 5.690 | ± 0.003(ran) ± 0.037(sys) |
| ω² Sco | 0.84 | 6.07 | ± 0.02(ran) ± 0.05(sys) | 5.743 | ± 0.004(ran) ± 0.038(sys) | 5.689 | ± 0.004(ran) ± 0.037(sys) |
| ε Sco | 1.16 | 6.07 | ± 0.02(ran) ± 0.04(sys) | 5.731 | ± 0.013(ran) ± 0.038(sys) | 5.683 | ± 0.006(ran) ± 0.037(sys) |
| α Sco A | 1.84 | 6.41 | ± 0.02(ran) ± 0.04(sys) | 5.735 | ± 0.006(ran) ± 0.046(sys) | 5.693 | ± 0.003(ran) ± 0.037(sys) |

Table 4 gives the final image scales and estimated uncertainties for each calibration method in our work. The results are averaged image scales for each technique, and the diffraction grating calibration is an average of the image scale determined across all stellar spectral types.

The percentage uncertainty clearly indicates that video drift calibration gives the best result. However, the size of the formal uncertainty for the video drift technique is possibly smaller as it results from a large ensemble of observations (over 300, see Paper 2) made through the observing session. Of the grating/filter methods, the accuracy of the technique falls off with larger filter bandpass. The Hα filter shows the best promise, however, as shown below, it results in a bias that makes use of the technique suspect.

*Table 4: The image scale determined by each calibration method.*

| Calibration Method | Image Scale px/arcsec | Uncertainty px/arcsec | No. Obs |
|---|---|---|---|
| Video Drift | 5.657 | ± 0.002(ran) | 310 |
| α Cen AB | 5.664 | ± 0.023(ran) | 40 |
| Grating/Hα (Grating) | 5.688 | ± 0.004(ran) ± 0.037(sys) | 60 |
| Grating/TR TriColour | 5.738 | ± 0.007(ran) ± 0.043(sys) | 60 |
| Grating/Wratten #25 | 6.11 | ± 0.02(ran) ± 0.06(sys) | 60 |
| Modelling (Grating/Hα) | 5.697 | − | |

## 6.2 Effect of Stellar Spectrum.

As stated, a grating calibration depends on the wavelength of the light passing through the system. For a broad-band filter, the colour of the star also affects the mean value of λ used in equation 1. Stars that are blue in colour move the mean λ towards lower values, and red stars move λ to higher values.

The colour of the star is best estimated by its Colour Index (CI) which is computed as B-V. The B-V values for the observed stars are in Tables 1 and 3 and have been accessed from the SIMBAD Astronomical Database.

Figure 5 is the image scale as determined using the 3 grating/filters combinations. The Wratten #25 filter is broad-band and allows a substantive contribution from the red end of the stellar spectrum. It is the most affected by the colour of the star.

Figure 6 also shows the trend of image scale with B-V values; here for the Hα filter (orange), TR TriColour filter (green). Because the band-pass is smaller for these filters the dependence of the image scale on star colour is less. The α Cen AB calibration (dotted blue), video drift calibration (dotted yellow), and Hα/diffraction modelling (dashed black) are also shown. There is no slope with α Cen AB or video drift calibration in Figure 6 as neither method relies strongly on the spectral type of the star. The Wratten #25 calibration was omitted to allow expansion of the vertical scale.

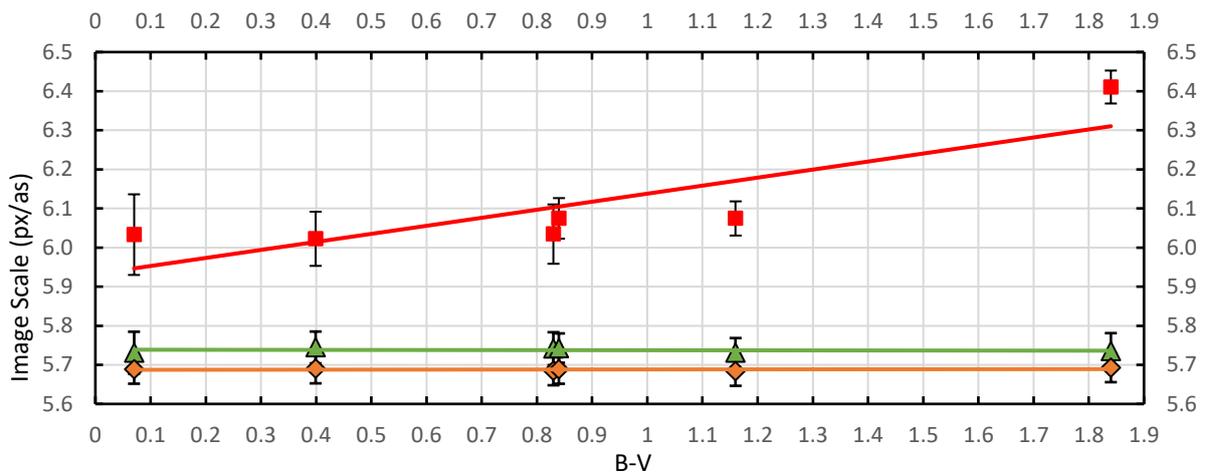

*Figure 5: Comparison of the image scale as determined using the Wratten #25 (shown red), the TR-TriColour (green) and Hα (orange) filters against star colour (B-V).*

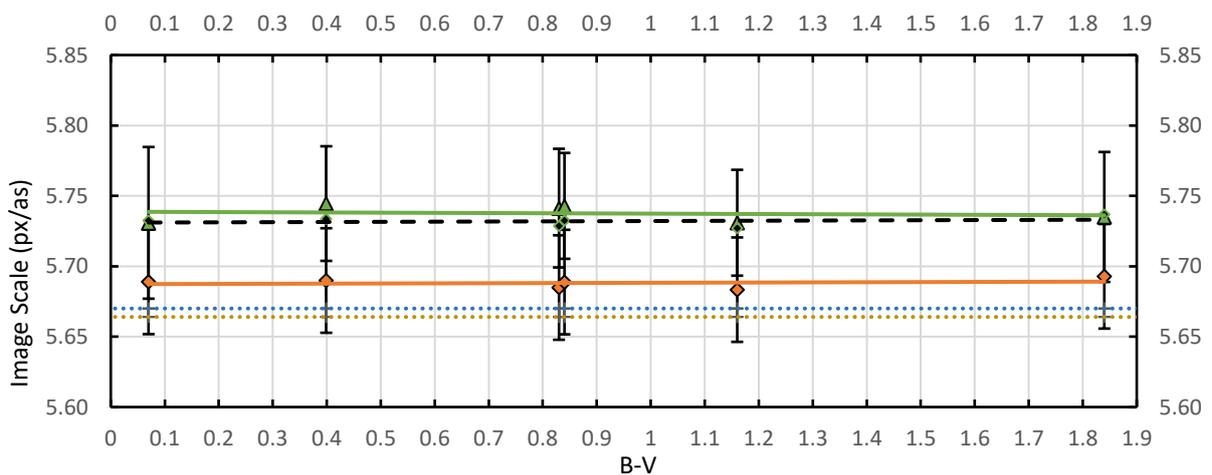

*Figure 6: Sub section of Figure 5. Comparison of the image scale as determined using the TR-TriColour (green) and Hα (orange) filters against star colour (B-V). The image scale for the Wratten #25 filter is not shown here. The α Cen AB calibration (dotted blue), video drift calibration (dotted yellow), and Hα/diffraction modelling (dashed black) are shown.*

## 7. Comparison of Calibration Methods.

### 7.1 Unexplained Grating Bias.

Figure 7 shows the separation of the 62 pairs presented in Paper 2, relative to the separation determined with the full aperture grating with the Hα filter. The separations were determined by (i) the α Cen AB calibration method (shown as ○), (ii) the video drift method (shown as △), (iii) extrapolated WDS historical data (shown as ×) and (iv) the extrapolated GAIA DR2 positions (shown as □). Here each difference is plotted against the separation.

It can be seen that the grating/Hα filter calibration produces, on average, separations larger than those of the other methods. The α Cen AB calibration shows the grating/Hα calibration to be $0.079 \pm 0.008$ arcsec larger. Similarly, the video drift, *Hist* and GAIA calibrations also indicate a bias of $0.108 \pm 0.019$, $0.089 \pm 0.023$, and $0.099 \pm 0.011$ arcsec respectively. Overall, the grating/Hα calibration results in separations that are ~0.1 arcsec larger.

The source of this discrepancy was not determined but two sources are thought to be responsible. First is smoke from the 2019/2020 Australian bushfires; the smoke at the time of observations was not noticeable. The smoke, through scattering, would shift the mean wavelength towards red leading to a bias in the grating/Hα image scale in the direction seen in this work. Secondly is the thermal expansion of the ABS plastic used to print the grating. The measurement of the grating spacing and slat width where done at a temperature ~20°C warmer than when observations were conducted. Calculations using typical literature values of the linear expansion coefficient for ABS plastic show that expansion may contribute ~50% of this bias.

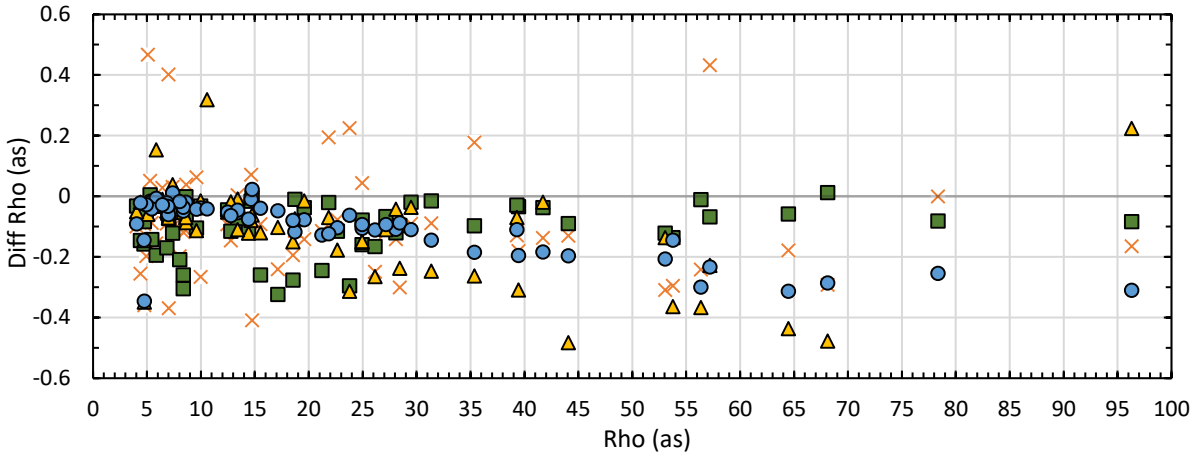

*Figure 7: The comparison of separations computed from different calibration methods. Hα − (Video Drift (△), α Cen (○), and Hist (×), and GAIA DR2 (□)).*

## 8. Discussion and Conclusion.

The purpose of this paper is to compare three methods of image scale calibration for the determination of measures of wide double stars. Table 4 gives the results of 3 methods, two of which have been used by us in our Papers 1 and 2. The third method is explored in this paper.

The calibration we obtained using the 3D printed diffraction grating proved difficult in procedure and accuracy; and, despite an inordinate amount of care, an unknown systematic bias occurred with all 3 grating/filter combinations. This was even the case for the narrow-band Hα filter which should have been relatively clear of bias resulting from a poorly defined wavelength (for equation 1).

The use of the grating and filter (by measuring fringe spacing) proved unsatisfactory for the broad-band filters, and the use of a narrow band Hα filter with the grating, resulted in image scales that differed from those obtained using α Cen AB reference pair calibration and the video drift method by 0.024 and 0.031 pixel/arcsec (px/arcsec) respectively. A more complete modelling of Fraunhofer diffraction of the Hα filter and grating produced a difference in image scale of 0.009 px/arcsec.

### 8.1 Preferred Image Scale Calibration Techniques.

The α Cen AB method requires the computation of the separation of α Cen AB on the date of observation. The advantage of this approach is that repeated observations can be made through the observing session – assuming that α Cen is high in the sky. To alleviate this issue – at least for southern and low northern hemisphere observers − we have in preparation a list of southern calibration pairs suitable for round-year observations of southern pairs (James *et al.*, in prep).

The video drift method is most convenient as no *a-priori* knowledge is needed besides the rotation rate of the Earth and the declination of the object. This technique requires the user to reset the position of the telescope multiple times for repeated observations needed for a reliable calibration, however, should not be a problem with a well-adjusted polar-aligned mount.

On the basis of this work we do not recommend the full aperture grating method calibration. Excluding operational difficulties, we found this approach to have a bias of ~0.1 arcseconds in separation values (relative to the α Cen AB and video drift method).

We conclude that multiple calibration with the video drift method the simplest and best way to undertake image scale calibration.

**Acknowledgements:**

We acknowledge the following sources:
- The University of New England (UNE) for use of the Kirby observatory and the *Raise3D pro2 3D printer* part of the School of Science and Technology.
- David Luckey for his expertise of the 3D printer and advice on improvements to the grating model.
- John Jarman for the construction of the camera/eyepiece flip box for the telescope.
- SIMBAD Astronomical Database, operated at CDS, Strasbourg, France, https://simbad.u-strasbg.fr/simbad/
- The Aladin sky atlas developed at CDS, Strasbourg Observatory, France, https://aladin.u-strasbg.fr/
- The Washington Double Star Catalog maintained by the UNSO. (WDS), https://ad.usno.navy.mil/wds
- "This research has made use of the VizieR catalogue access tool, CDS, Strasbourg, France (DOI : 10.26093/cds/vizier). The original description of the VizieR service was published in 2000, A&AS 143, 23.

- All-sky Compiled Catalogue of 2.5 million stars, 3rd version (ASCC), http://vizier.u-strasbg.fr/viz-bin/VizieR-3?-source=I/280B/ascc
- The GAIA Catalogue (GAIA DR2, GAIA collaboration, 2018), from VizieR, http://vizier.u-strasbg.fr/viz-bin/VizieR-3?-source=I/345/gaia2
- The 3D modelling software Google *SketchUp* developed by Timble Inc, https://www.sketchup.com/